# M-Learning: A New Paradigm of Learning Mathematics in Malaysia

Saipunidzam Mahamad[1], Mohammad Noor Ibrahim[2] and Shakirah Mohd Taib[3]

Department of Computer and Information Sciences, Universiti Teknologi PETRONAS,
Bandar Seri Iskandar, Perak, MALAYSIA

[1]saipunidzam_mahamad@petronas.com.my, [2]mnoor_ibrahim@petronas.com.my,
[3]shakita@petronas.com.my

## ABSTRACT

*M-Learning is a new learning paradigm of the new social structure with mobile and wireless technologies. Smart school is one of the four flagship applications for Multimedia Super Corridor (MSC) under Malaysian government initiative to improve education standard in the country. With the advances of mobile devices technologies, mobile learning could help the government in realizing the initiative. This paper discusses the prospect of implementing mobile learning for primary school students. It indicates significant and challenges and analysis of user perceptions on potential mobile applications through a survey done in primary school context. The authors propose the m-Learning for mathematics by allowing the extension of technology in the traditional classroom in term of learning and teaching.*

## 1. INTRODUCTION

M-learning introduces a new learning environment due to the emergence of mobile and wireless technologies. It offers a new way to deliver learning objects into our daily life. This could be done by developing learning materials in small and consumable byte of format to its delivery medium. Liang Ting [1] stated that there are various mobile communication mechanisms that support m-Learning such as voice communication, access of learning portal on the internet and learning through SMS. This clearly shows that m-Learning could be interactive with the convergence of audio, web and mobile technologies in one package.

The government of Malaysia has established MSC (Multimedia Super Corridor) to spearhead the four innovative flagship applications where one of them is to implement smart schools application. However, there are still a large number of schools and educational institutions do not use the available technologies even though those schools are equipped with high-technology equipments to support a new learning approach. The smart schools will use new curriculum for Malay language, English, Mathematic and Science with objectives of smart school to improve the learning quality, training, school organization and student presentation. In addition, the learner will develop the critical thinking, learning activities and creativities.

Mathematic has come to our apprehension as its nature has become a core subject must be taken in primary school examination in Malaysia named UPSR (Ujian Pencapaian Sekolah Rendah or Primary Schooling Achivement Tests). Besides, the Malaysian Examination Board also reported that more than 35% of primary school student have fail the course [2]. Teaching and learning mathematics have faced many challenges worldwide that either in delivery methods and student's participations [3]. There are four main problems in the current learning systems in Malaysia that inspires us to work on this project. The problems mainly dwell around the current learning environments, which is too rigid that limit the students' potential in their learning processes. The problems are:

- heaviness of the textbooks used.
- student and teachers having hectic daily schedule.
- unnecessary and confusing e-learning function for primary school students.
- high cost in commercialize system and desktop use.

According to a study done by Chong et al [3], presentation apparatus and courseware have been used extensively in teaching and learning the subject. The teachers still need more time to learn other advanced applications and how to integrate ICT tools in their lesson. One of the major barriers faced by mathematics teachers in Malaysia during integration is insufficient of time on their timetable for projects relating to ICT. This problem could be solved with portable communication devices as their mobility allows teaching and learning to be more flexible and provides new opportunities for interaction [4].

In line with the knowledge work force emphasized by the government, it is crucial for everybody to instill the lifelong learning philosophy. M-Learning could be the best platform for everybody, at any ages to keep on learning. Short and small courses for distance learning could employ m-Learning in transferring the knowledge and providing an interactive yet valuable experience for the learners everywhere at any time.

This paper presents the Malaysian prospect of implementing M-Learning that includes significant and challenges of m-Learning, and user perceptions.  This new learning approach is to make variety of choice and utilizes the advancement of today technology that embedded with the current way of learning. The framework of M-learning is focusing on learning mathematic in primary schools in Malaysia. It includes mobile quizzes and progress tracking where students to attempt the mobile quizzes online, and teachers as well as parents to monitor the progress of the students.

## 2. MOTIVATION FACTORS

Researchers [1, 4, 5] had agreed that M-learning is a new trend of learning paradigm with the emergence of mobile and wireless technologies usage among the learners. Even though some of the countries, especially the developing countries are still in the first phase or perhaps in the research and development  phase in implementing this type of learning environment, Kyun Baek and Uk Cheong [6] and Barker, Krull and Mallinson [7] had proved that developing countries as well will soon catch up with this new learning paradigm. This shows that this new learning paradigm will evolve mobile devices with the rapid usage and ownership among the users. The usage of those devices, as an example, the hand phones could be extended to create the new learning environment to the owner instead of using it for the sake of making or receiving a call and SMS (Short Massaging System).

M-learning will be a successful trend currently and in the future because currently, PDA and tablet PCs are more popular among users because of several factors [8]. One of the factor is that the mobile devices are inexpensive compared to PC. Besides that, the devices are mobile, durable and convenient for the worker on the go and lastly, the familiarity among younger users makes them an attractive mechanism for incorporating M-learning into the curricular. The evolution of learning paradigm from traditional classroom based learning and electronic learning had brought out the new learning paradigm based on mobile devices which is known as m-Learning [9].

A mobile technology have become pervasive, many researchers have questioned whether they can enhance learning experiences. Arguably, it could be thought that m-Learning is an utilization of mobile devices in e-Learning environment that picture a different skill on learning. This is because e-Learning and m-Learning are totally different in terms of its mobility and interaction among the students, as well as the teachers.

The development of e-Learning is not intended to replace the classroom or PC based (e-Learning) learning content, but to strengthen and harmonize overall learning strategy. Meanwhile, m-Learning offers another way to deliver content and to embed learning into daily life by developing learning materials in small and consumable byte of format which can be delivered through wireless network [1].

## 3. M-LEARNING IN VIEW

### 3.1. Challenges of Implementing M-Learning

According to Barker, Krull and Mallinson [7], the challenges of implementing m-Learning are device limitations, issues on instructional, training, safety, security, and maintenance, and the implementation cost.

In term of limitations of the devices, the devices being used in implementing this type of learning environment might not give the same layout or interface of the content as the emulator. This had been agreed by Black and Hawkes [8] where they had stated that the different output or layout of the content might be differed to the one that is appear in the emulator. This problem should be considered as technical problem since several adjustment to the codes or settings of the devices should be made in order to ensure the same outcome appear to the devices as well as the emulators. The example of limitations of the devices also includes the mobile device capability in executing the program used in m-Learning, whether it manages to give response on time or not. Besides, the mobile phone technologies nowadays are also being proprietary products of its developer. Variety of screen size and functionalities make difficult for the developer to design for all.

Pedagogical is a learning model that focuses on teachers who control the learning process. According to Keough [10], a new learning philosophy which is called mobigogy that integrates pedagogy (teacher centered learning) and andragogy (learner oriented) should be adapted to M-Learning environment. This is because; m-Learning is no longer being controlled by teachers fully in the classroom, but also makes it possible for learners to learn anything that they want to learn, at any time and at any places even though outside the school area.

The hand held devices owned by school are prone to damage, lost, misuse and other problems were strengthening many school from allowing learner to bring home the devices. This safety and security issues will eventually made the learners who are from low-income family have the difficulty in owning the devices to facilitate them in M-Learning environment. In terms of safety, the usage of the wireless internet without any supervision might lead to the learners to join negative group, which might threaten the learners' safety.

In every new technology, training and support are very important and m-Learning implementers need to take extra consideration about it. Learners and teachers, as well as the parents should be taught a lesson on devices function to fully utilize the m-Learning environment. Any problems such as troubleshooting services should be ready anywhere and at anytime if the stakeholders, in this case the learners, teachers and parents, faced any problems in using the technologies. Even though these are quite a dawn tasks, it should be taken extra considerations to ensure the stakeholders' could benefits fully from this new learning environment.

The cost of the technologies and infrastructures in implementing m-Learning environment without any doubt will be very high. The cost of the mobile devices itself still being considered as expensive since there are prove that the devices price is reducing. The term technologies refers to the programs or systems used to develop mobile based system while the term infrastructure refers to the wireless network hardware and devices used to create the framework of mobile communication.

According to Liang Ting [1], the challenges of implementing m-Learning are in terms of location oriented learning content, cognitive effectiveness of information and the effect of instant interaction upon learning interaction. In implementing M-Learning, learning content should be designed to adapt to learners profile and personal needs such as location. The architecture of the mobile devices itself is small and this eventually limiting the text display in supporting the learning process. Instant communication in mobile network is very important in making the learning process more exciting and this is based on the location and the response time. Slow response time will demoralize the learners' learning process. Kyun Baek and Uk Cheong [6] agreed that the requirements for future mobile learning implementations are m-Learning can only be accepted only based on acceptance of the beliefs that learning can happen everywhere and at any time when there is a need for it, the adaptive learning strategies implies that learning should be flexible and specific to the learner's style, the establishment of infrastructure and the standardization of the contents.

Another important issue to be considered is the different specifications and attributes of mobile devices. For example if a mobile device that only supports GIF format will incorrectly display any learning object that created using other format. Therefore, Zhao and Okamato [11] conclude that learning content should be matched with or adaptable to the varying capabilities of mobile devices. Black and Hawkes [8] stated the challenges of implementing m-Learning using J2ME technology are the lack of documentation on J2ME itself, different interface from emulator and the actual device, slow file retrieval from server to client and because of security and reliability issues. Their statement refers to the technical challenges in employing M-Learning with the inclusion of the technologies and infrastructures such as Java 2 Micro Edition (J2ME) and the wireless communication infrastructure.

Based from the research arguments, we can divide the challenges in implementing m-Learning environment into two aspects which are the management aspects, which includes the pedagogical, training and support issues and the technologies and infrastructures aspects, which includes the cost, compatibility and limitation of the elements.

## 3.2. Significance of M-Learning

The future of M-Learning seems bright since new types of wireless communications services such as wireless collaboration will be widely available soon with wireless and wired Internet services are popularized. The development of mobile semiconductors such as flash memory will even make the mobile devices smaller and this will give impact on the usage group of wireless services where students use wireless internet more than graduates [6].

It is human nature to be attracted to any appealing things and same goes to school students as well. Students at any age are attracted to fun and new activities using new devices. With the technology new learning concept could be applied and it is practical to implement this learning environment for that age group using the ever appealing mobile devices. They could be tempted to involve in the learning process dynamically if the medium is appealing to them in term of the devices' weight itself which is multiple times lighter than their mathematics textbook.

Besides, everybody likes to stay connected with their devices at any time and at any place in order to ease their work. The confusion that happened by using the unnecessary functions on the current E-learning system can be solve by implementing learning processes using mobile devices. The limited amount of memory in the mobile devices had urged the mobile content developers to include minimum amount and easy to use functions in their product. The nature of m-Learning which includes only necessary functions could make the students feel easy to use the system to its maximum to enhance their learning processes.

The hectic schedule faced by both students and the teachers could demoralize them to participate in the learning processes dynamically. The implementation of m-Learning with automated progress tracking function using graph could solve the problem of time constraints faced by both students and teachers. The students, as well as the teachers can easily monitor the progress after they had attempted the quizzes using the mobile devices. The progress could easily be monitored without any intervention and use of paper. The students and the teachers can take necessary actions to improve the students' performance after analyzing the automated graph displayed on the mobile devices. Mobile devices enable student to explore and experiment the concepts that been learnt while teacher are able to control the difficulty levels which appropriate to their student capabilites [12].

 Personal m-devices are used instead of public desktops, that need to be funded by the schools themselves, had make M-Learning more appealing in terms of cost issue. The implementation of M-Learning using Open Source Software (OSS) indirectly will reduce the implementation cost. PHP and MySQL, which are the examples of Open Source technologies, are the main technologies could be used in the development of such system.

## 3.3. Benefit of Mobile Technology to be used in Learning

The benefits of mobile learning, as stated by Kyun Baek and Uk Cheong [6] are the possibility of implementing ubiquitous learning environment, the possibility of lifelong learning and the possibility of education through entertainment 'edutainment'. Ubiquitous, according to Oxford Advance Learner's Dictionary means seeming to be everywhere or in several places at the same time. With the development of M-learning environment, one will be possible to learn everywhere at any time. A student who is on a holiday could still read the lecture notes and doing the exercises using his or her mobile devices. Collaborative learning could be possible even if the learner is not in the classroom.

Games had always been the favorite activity of the youngsters. Even mobile technology companies such as Nokia and Motorola had equipped their products, which is the hand phones with various mobile games. If learning materials could be integrated with that type of entertainment, the youngsters will be more eager to learn and understand the contents of their learning. The technicality and increasing functionality of mobile devices such as hand phones and PDA's could attract learners in learning efficiently.

According to Stead [9], Learning through mobile application could remove the barrier in Information Technology (IT). It means that the users of M-learning will be more confident in dealing and working with IT technology such as mobile phones and PDA's. Currently, there are lots of people, particularly with socially disadvantaged groups had no confidence in IT at all. Thus, with the implementation of M-learning at young age could overcome this problem which faced mostly by developing countries.

There are several issues of implementing the MobileMath application such as were mobility in a learning environment and the impact of progress monitoring in learning environment. The mobility allow users, particularly the students a freedom to learn anywhere and at anytime at their own pace by using their personal mobile devices. This proposed framework enables mobility in learning by implementing learning content through mobile devices. This eventually had solved the problem of the students that are not interested to do academic exercises regularly because of the heaviness of the materials, especially the books.

Besides the issue of mobility, learning processes could be enhanced with the implementation of progress monitoring in the learning system itself. Progress monitoring can be depicted using graphs. Analyzing progress will be much easier if the data could be represented in a graphical manner. This project clearly champions the usage of progress monitoring functions in the m-learning system. The students could view their performance on the quizzes that they had attempted which is generated by a graph. The students can know their strengths and weaknesses on the subject and make necessary actions about it. The burdens of the teachers in analyzing their students' performance manually are reduced with the help of the computer generated graph instead of using any papers.

The implementation of M-learning for primary school in Malaysia by using an open source technology may prove the potential of a new mobile learning environment. The developed model named MobileMath [12] focuses on learning mathematics had allowed the learners to do a lesson, quiz, tests and performance tracking with automated graph. The development has shows significant contribution and improvement from Malaysian perspective on education environment. The general architecture of MobileMath is shown in Fig 1 that explains all of the conceptualize data gathered before this were put into actual physical model. It's illustrated the levels of user access includes Students, Teacher and Administrators which have different role of permission in accessing the content and functionalities. Each of the users has to go through an authorization function before proceed to the individual function such as lesson, difficulties setting and progress tracking.

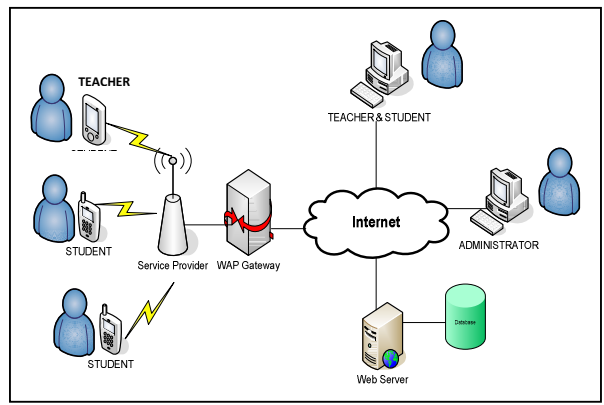

Figure 1: Mobile Math architecture

# 4. DISCUSSION

A survey has been conducted to investigate the use of mobile devices and to determine if primary school students were ready for mobile learning. 98 students from various primary schools volunteered to participate in the survey.

Figure 2 shows the mobile devices that currently owned by the students that participated in the survey. In general, most of the students already have experience using the mobile devices. The survey shows, from the 98 respondents, 46% owned mobile phone. Based on activities engaged in by students shown in Figure 3, it shows that they already participated in several of communication activities. About 115 of the responds engaged in two popular activities; sending and receiving SMS as well as downloading and playing phone game. The overall result has indicated their readiness for m-Learning.

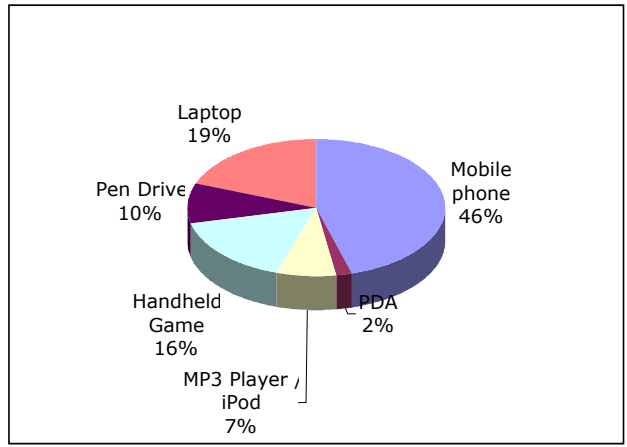

Figure 2: Mobile devices owned by primary school students

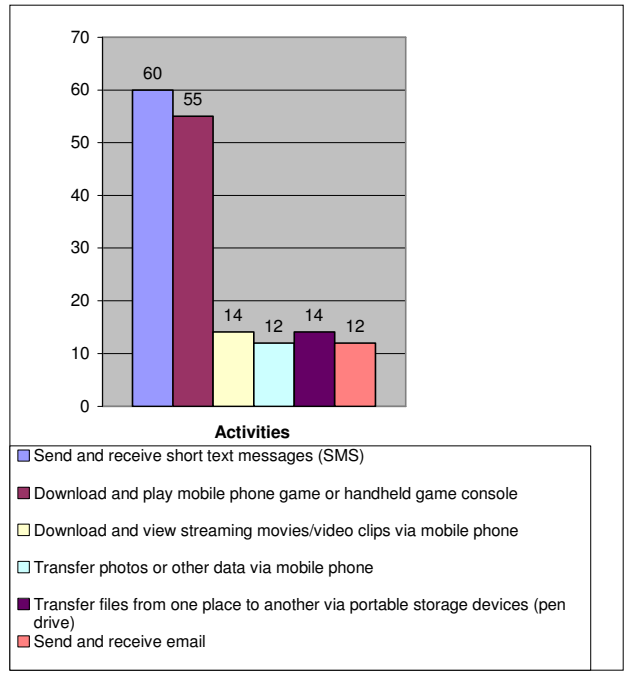

Figure 3: Communication Activities

Figure 4 illustrates that more than 50% of the respondents are ready to start m-Learning, 12% are undecided and another 31% are unwilling to learn with the help of mobile devices. M-Learning might not positively considered by some of the students because of the lack of knowledge about the devices.

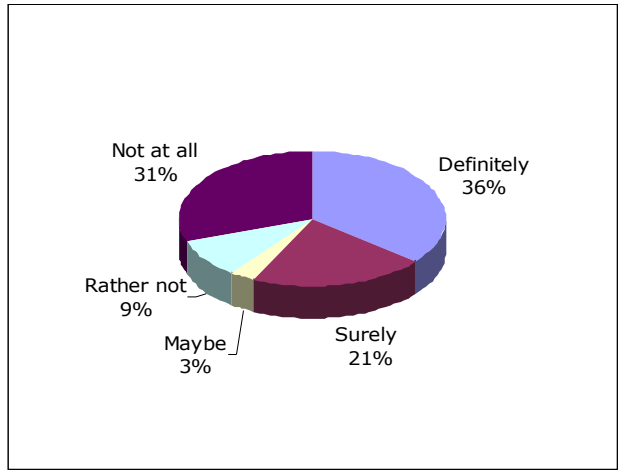

Figure 4: m-Learning readiness

Therefore it is recommended to start implementing m-Learning by enabling a simple format of content and information to the student that can accessible by mobile phone. Barker, Krull and Mallinson [7] has suggested the development of it should start with fully information and theory courses since it is much easier to suit with the m-Learning environment. As an example, subject like Mathematics which requires objective-type-questions is suitable to be implemented in M-Learning environment since the subject require only thorough understanding of certain basic mathematical concepts before attempting to answer the objective-type-questions.

Figure 5 illustrates the distribution of feedbacks to the question which types of features are appropriate to be included for M-Learning setting. Online notes and assessment are two favorite features viewed as most appropriate by the students.

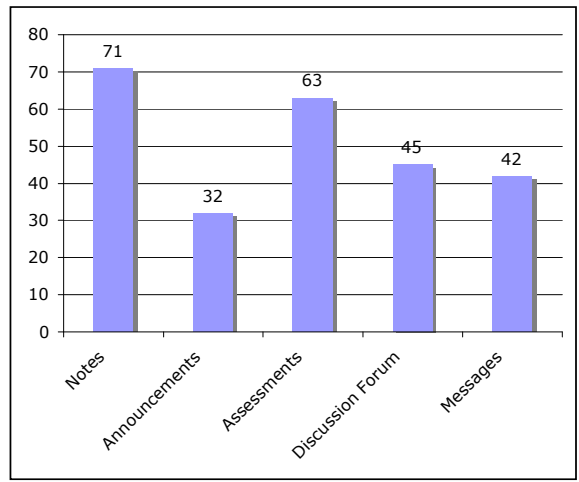

Figure 5: Recommended m-Learning System Features

One of the six major types of learning is behaviorism. M-Learning can be adapted to this type of learning by providing quick feedback through mobile devices [14]. From the survey conducted, 93 of the respondents agree that it would be good if they can have a quiz and view their performance while they are not in the class.

In spite of content and instructional design, another challenge of implementing m-Learning environment is the current technologies such as small mobile phone screen, limited memory and standardization of the operating system. Network connection problem also creates slow file retrieval from server to client. In the case of the mobile devices, it could be overcome by implementing such assessment such as the objective-type-questions [7, 9]. It is because, this type of processing requires only small number of computations compared to essay or subjective-type-questions which require high amount of computation which eventually make the response from the server slow.

## 5. CONCLUSION

Learning environment in Malaysia has been continuously revised from time to time. The Ministry of Education, Malaysia has been given the responsibility to enhance and develop the level of educations in Malaysia. This shows that the government really pays a lot of attention to the level of educations in Malaysia The Education Technology Division of Ministry of Education in 2006 had mentioned that one of their objectives is to increase the contact hours in the learning process between the students and the computers. These objectives could be achieved with the implementation of m-Learning. Since its focus on the usage of the students' own personal devices instead of the public PCs in the schools.

The advancement in the field of Computer and Information Technology had broadened the horizon in the environment of education. From traditional based learning environment to electronic learning (e-learning) environment, then, the new way of learning, known as mobile learning (m-Learning) had been implemented in developed countries such as the USA, UK and Japan. The content of m-Learning could be more appealing to the students since it is the new concept and a new way of learning. M-Learning could be implemented in Malaysia in the future. With the good framework being designed and the support of various organizations, the m-Learning environment could be realized successfully. The result of the survey shows that the mobile phones can be useful in learning mathematics as most of primary school students already use them through many communication activities. Mathematics teachers should start implementing the M-Learning to allow students to independently explore the lesson taught with flexible access to the content and construct the effective teaching environment.

## ACKNOWLEDGEMENTS

The author would like to acknowledge the support and contribution of this work to Mohamad Izzriq Ab Malek Foad.

**Authors**

**Saipunidzam Mahamad** was born in Perak, MALAYSIA on October 23, 1976. This author was awarded the Diploma in Computer Science on 1997 and the Bachelor Degree in Computer Science in 1999 from University Technology of Malaysia. Later, the author pursued his master degree in Computer and Information Science at the University of South Australia and was awarded the Master Degree in 2001. The author is a lecturer since January 2002 and is currently attached with University Technology PETRONAS (UTP). The author started with UTP as a Trainee Lecturer in July 2000. Apart from teaching, he is involved in Software Engineering and E-Commerce research group and does various administrative works. The author has been appointed as UTP's E-learning Manager on Apr 2006 - Sept 2008. This major contribution is to manage and monitor the new e-learning implementation at UTP using Moodle software. The author research interest include Computer System, E-Learning, Software Engineering and Intelligent Systems

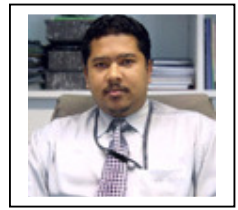

Mohammad Noor graduated with a Degree in Computer Science from Universiti Pertanian Malaysia in 1996. He started his career as Trainee Lecturer at Universiti Teknologi PETRONAS (UTP) in 1997 and later pursues his Master degree at Loughborough University. He has been awarded Masters of Science in Multimedia and Internet Computing in 2001 and has been appointed as Lecturer in Computer and Information Sciences Department in 2001. His research interest includes Mobile Application, Mobile Learning, Agile Software Development and Software Engineering.

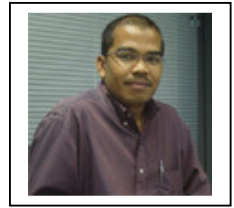

Shakirah Mohd Taib obtained her Master degree in Computing in 2004 from University of Tasmania, Australia. Since 2004 she is a lecturer in the Computer and Information Sciences Department, Universiti Teknologi PETRONAS. Her fields of interest are Knowledge Management, Artificial Intelligence, Web Mining and Decisions Support Systems. She has published technical papers in International, National journals and conferences.

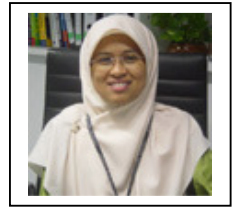